\documentclass[aps,prapply,twocolumn,footinbib,superscriptaddress,longbibliography,amsmath,amssymb]{revtex4-1}
\usepackage{graphicx}
\usepackage{subfigure}
\usepackage{epsfig}
\usepackage{dcolumn}
\usepackage{bm}
\usepackage{times}
\usepackage{amsmath}
\usepackage{float}
\usepackage{multirow}
\usepackage[normalem]{ulem}
\usepackage{color}

\usepackage{pifont}
\begin{document}
	\title{Twisting the near-field radiative heat switch in hyperbolic antiferromagnets}
	\author{Yuanyang Du}
	\affiliation{Center for Phononics and Thermal Energy Science, China-EU Joint Lab on
		Nanophononics, Shanghai Key Laboratory of Special Artificial Microstructure Materials and
		Technology, School of Physics Science and Engineering, Tongji University, 200092 Shanghai,
		China}
	\author{Jiebin Peng}
	\email{Email: jiebinxonics@tongji.edu.cn}
	\affiliation{Center for Phononics and Thermal Energy Science, China-EU Joint Lab on
		Nanophononics, Shanghai Key Laboratory of Special Artificial Microstructure Materials and
		Technology, School of Physics Science and Engineering, Tongji University, 200092 Shanghai,
		China}
	\author{Zhong Shi}
	\affiliation{Center for Phononics and Thermal Energy Science, China-EU Joint Lab on
		Nanophononics, Shanghai Key Laboratory of Special Artificial Microstructure Materials and
		Technology, School of Physics Science and Engineering, Tongji University, 200092 Shanghai,
		China}
	\author{Jie Ren}
	\email{Email: xonics@tongji.edu.cn}
	\affiliation{Center for Phononics and Thermal Energy Science, China-EU Joint Lab on
		Nanophononics, Shanghai Key Laboratory of Special Artificial Microstructure Materials and
		Technology, School of Physics Science and Engineering, Tongji University, 200092 Shanghai,
		China}
	\begin{abstract}
		We study the twisted control of the near-field radiative heat transfer between two hyperbolic antiferromagnetic insulators under external magnetic fields. We show that the near-field heat flux can be affected by both the twist angle $\theta$ and the magnitude of the applied magnetic field with different broken symmetries. Irrespective of twist angle, the external magnetic field causes the radiative heat flux to change nonmonotonically, and the minimum heat flux can be found with the magnetic fields of about 1.5 T. Such nonmonotonic behavior is due to the fact that the magnetic field can radically change the nature of the magnon polaritons with time reversal symmetry breaking. The field not only affects the topological structure of surface magnon polaritons, but also induces the volume magnon polaritons that progressively dominate the heat transfer as the field increases. We further propose a twist-induced thermal switch device with inversion symmetry breaking, which can severely regulate radiative heat flux through different magnetic fields. Our findings account for a characteristic modulation of radiative heat transfer with implications for applications in dynamic thermal management.
	\end{abstract}
	\maketitle
	\bigskip
	\section{Introduction}
	Thermal radiation is one of the most common phenomena, with properties described by the well-known Planck's Law. Presently, the investigation of radiative heat transfer between closely spaced objects is receiving much attentions. The traditional Stefan-Boltzmann law can be broken at nanoscale, as coupling of evanescent waves provides extra radiation channels for heat transfer\cite{Biehs-2013,Cahill-2014,Polder-1971,Volokitin-2007}, which has been verified in various experiments exploring different materials, geometrical shapes and gaps ranging from micrometers to a few nanometers\cite{Domoto-1970}.
	These experiments have also triggered off the hope that near-field radiative heat transfer (NFRHT) could have an impact in different technologies such as date carrier storage\cite{Pendry-1999,Challener-2009,Stipe-2010}, coherent thermal sources\cite{Carminati-1999,Greffet-2002}, scanning thermal microscopy\cite{Yannick-2006,Kittel-2008,Jones-2013}, active noncontact thermal management\cite{Zwol-2012,Otey-2010,Abdallah-2014}, thermophotovoltaics\cite{Messina-2013,Park-2008} and other energy conversion devices\cite{Schwede-2010}.
	
	In recent years, one of the significant research lines in the territory of radiative heat transfer is searching for materials where the NFRHT can be greatly enhanced. Up to now, the largest NFRHT enhancements have been reported for polar dielectrics, in which the NFRHT is dominated by surface phonon polaritons (SPhPs)\cite{Mulet-2002,Iizuka-2015}. Similar enhancements have been predicted and observed in doped semiconductors due to surface plasmon polaritons (SPPs)\cite{Jiawei-2013,Lim-2015}. Subsequently, several van der Waals heterostructures can support surface phonon-plasmon polaritons (SPPPs), which also enhance the NFRHT\cite{Brar-2014,Kumar-2015,Dai-2015}. Apart from these, surface magnon polaritons (SMPs) which can be excited in magnetic media play a significant role in NFRHT due to its striking hyperbolic behavior\cite{Jiebin-2021,Rair-2019}.
	
	Another key issue in the territory of radiative heat transfer is the active control and modulation of NFRHT. Several strategies have been proposed, such as applying an electric field to phase-change materials\cite{Zwol-2011} or ferroelectric materials\cite{Huang-2014}, applying a magnetic field to magneto-optical materials\cite{Moncada-2015,Latella-2017} or magnetic Weyl semimetals\cite{Gaomin-2021,Fan-2020,Zhao-2020}. Another active control strategy to utilize the rotational degree of freedom\cite{Biehs-2011,MingJian-2020}, similar twist-induced concepts have been demonstrated in two-dimensional materials\cite{Yuan-2018,Yuan-2020} and photonics\cite{Sunku-2018,Guangwei-2020}, this control strategy is called twisting method.
	
	In this work, we present a theoretical analysis of the impact of an external magnetic field in the radiative heat transfer between two parallel plates made of twisted antiferromagnetic insulator (AFMI). We find that the presence of a magnetic field makes the radiative heat flux change nonmonotonically and we show that the reduction can be large as 25$\%$ for the fields of about 1.5 T. This phenomenon due to the fact that the magnetic field not only strongly modifies the width of the band where SMPs can be excited that normally dominate the NFRHT, but also generates the nonreciprocal volume magnon polaritons (VMPs) via time reversal symmetry breaking, which can dominate the heat transfer as the field increases. Apart from these, we utilize the nonreciprocity of VMPs to get a twist induced thermal switch. To enhance twist induced thermal switch ratio, we actively select the heat transfer channels sensitive to twist through different applied magnetic fields. Thus, our work offers an effective strategy to regulate NFRHT between the antiferromagnets and provides an attractive recipe to design device with heigh adjustable characteristics.
	
	\section{Theoretical Formalism for near-field radiative heat transfer}
	
	\begin{figure*}
		\centering	
		\includegraphics[width=\textwidth]{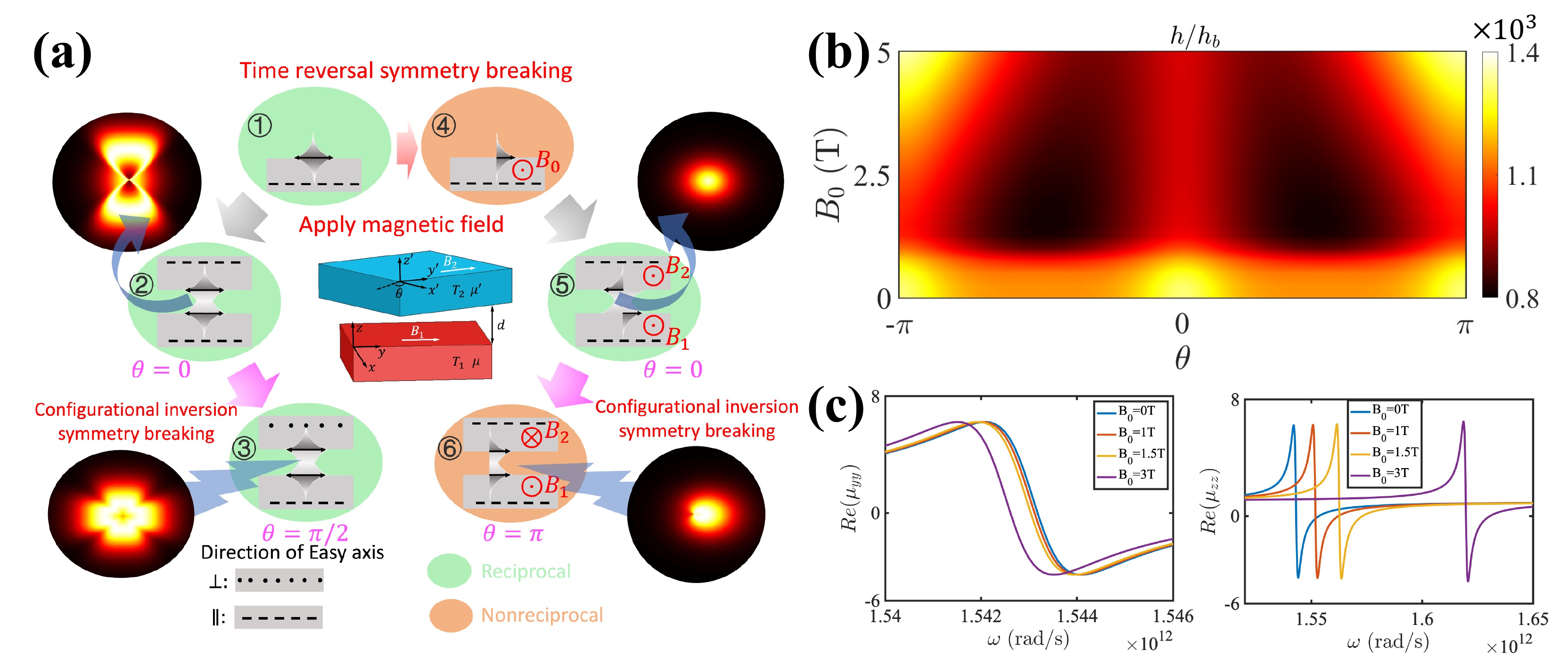} \\
		\caption{(a) Schematic of the near-field radiative heat transfer between two semi-infinite AFMIs, the thickness of vacuum gap is $d$. The twist angle $\theta$  is defined by the anticlock wise rotation of ${x}^{\prime}{y}^{\prime}{z}^{\prime}$ coordinate system with respect to ${x}{y}{z}$ coordinate system. The energy transmission coefficients for wave vector space reflect the symmetry of electromagnetic mode. Note that applying magnetic field and twisting broke the time reversal symmetry and configuration inversion symmetry. (b) The heat transfer coefficient varying with magnetic field at different twist angle. (c) Real part of $y$ and $z$ direction component of permeability tensor of AFMI.}
		\label{fig1}
	\end{figure*}
	
	As schematically depicted in Fig. \ref{fig1}(a), the system consists of two twisted AFMI slabs in the presence of magnetic field. The structures are separated by a vacuum gap of width $d$ and held at temperatures $T_{1(2)}$, where the magnetic field $B_{1}$($B_{2}$) applied along $y(y^{'})$ direction as the relative twist angle expressed $\theta$. From fluctuational electrodynamics, the heat transfer coefficient (HTC) between two AFMI slabs for a temperature $T$  is given by the following Landauer-like expression\cite{Biehs-2013}:
	\begin{align}
		h(T,\theta,B)=\int_{0}^{\infty}\frac{\partial\Theta(\omega,T)}{\partial T} \Phi(\omega,\theta,B)d\omega,
	\end{align}
	where $\Theta(\omega,T)={\hbar \omega}/ {[e^{(\hbar\omega/k_{B}T)}-1]}$ represents the average photon energy of Planck oscillators at an angular frequency $\omega$ and temperature $T$, and $\Phi(\omega,\theta,B)=\int_{-\infty}^{\infty}\int_{-\infty}^{\infty}\xi(\omega,k_{x},k_{y})/8{\pi}^{3} dk_{x} dk_{y}$, where  $k=\sqrt{k_{x}^{2}+k_{y}^{2}}$ is the wave vector parallel to the surface planes and
	$\xi (\omega,k_{x},k_{y})$ with twist angle $\theta$ characterizes the energy transmission coefficient that describes the probability of photon excited by thermal energy, which can be expressed as\cite{Biehs-2011}:
	\begin{align}
		{\cal \xi }= \left\{
		\begin{aligned}
			&{{\rm Tr}[({\bf I}-{\bf R}^\dagger_2{\bf R}_{2}) {\bf D} ({\bf I}-{\bf R}_1{\bf R}^\dagger_1) {\bf
					D}^\dagger] }, & k < \omega/c , \\
			&{{\rm Tr}[({\bf R}^\dagger_2 -{\bf R}_2) {\bf D} ({\bf R}_1 - {\bf R}^\dagger_1) {\bf D}^\dagger] }
			e^{-2|k_{z}|d}, & k > \omega/c,
		\end{aligned}
		\right.
	\end{align}
	where $k_{z}=\sqrt{k_{0}^{2}-k^{2}}$ represents the wave vector vertical to the surface planes in the vacuum gap, and $\dagger$ signifies the complex conjugate. $\bf I$ is a $2\times 2$ unit matrix and the  $2\times2$ matrix ${\bf D}$ defined as ${\bf D}=({\bf I}-{\bf R}_1 {\bf R}_2 e^{2ik_{z} d})^{-1}$, which describes the Fabry-perot-like denominator. The $2\times 2$ matrix ${\bf R}_{a}$ is in the reflection coefficient tensor of different polarizations.
	
	To clarify the enhancement of the heat transfer channel coupling degree and simplify analysis, we first consider the situation that the external magnetic fields are same value, i.e., $B_{1}=B_{2}=B_{0}$. In this case, the permeability tensor of the bottom AFMI slab has the form\cite{Almeida-1988}:
	\begin{align}
		\mu =
		\begin{bmatrix}
			\mu_{xx} & 0 & \mu_{xz} \\  0 & \mu_{yy} & 0 \\ \mu_{zx} & 0 & \mu_{zz}
		\end{bmatrix} ,
	\end{align}
	where
	\begin{align}
		\label{tensor}
		\begin{split}
			&\mu_{xx}=1+\frac {2\mu_{0} \gamma^{2} M_{S} B_{0} \sin{\alpha}}{{\omega_{{\perp}}}^{2} - {(\omega+i\Gamma)}^{2}},\\
			&\mu_{yy}=1+\frac {2\mu_{0} \gamma^{2} M_{S} B_{A} {\cos^{2}{a}} }{{\omega_{{\parallel}}}^{2} - {(\omega+i\Gamma)}^{2}},\\
			&\mu_{zz}=1+\frac {2\mu_{0} \gamma^{2} M_{S} (B_{0} \sin{\alpha}+B_{A} \cos{2\alpha})}{{\omega_{{\perp}}}^{2} - {(\omega+i\Gamma)}^{2}},\\
			&\mu_{xz}=-\mu_{zx}=-i\frac{2\mu_0 \gamma (  M_{S} (\omega+i\Gamma)\sin{\alpha})}{ {\omega_{{\perp}}}^{2} - {(\omega+i\Gamma)}^{2} },
		\end{split}
	\end{align}
	where, $B_{A}$ measures the anisotropy force and $B_{E}$ is the exchange force that is exerted on each ion by the ions forming the other sublattice. The sublattice magnetization $M_{s}$ will point  predominantly parallel and antiparallel to a preferred axis, which along the $x$-direction. Then, $\gamma$ and $\Gamma$ are gyromagnetic ratio and damping parameters, respectively\cite{Johnson-1959}. Moreover, $\omega_{\parallel}$ and $\omega_{\perp}$ are the components of the resonance parallel and perpendicular to magnetic field and depend on the value of magnetic field , these can be express as
	\begin{align}
		\begin{split}
			&{\omega_{{\parallel}}}^{2}={\omega_0}^{2} {\cos^{2}{\alpha}},\\
			& {\omega_{{\perp}}}^{2}={{\omega_0}^{2} {\cos^{2}{\alpha}} + 2 \gamma^2 B_0 B_E \sin{\alpha} },
		\end{split}
	\end{align}
	where $\omega_{0}={\gamma}(2B_{A}B_{E}+{B_{A}}^{2})^{\frac{1}{2}}$ represents the resonance frequency in the absence of magnetic field and the spins cant at an angle $\alpha=\arcsin{\frac{B_{0}}{B_{A}+2B_{E}}}$ in the $x-y$ plane\cite{Almeida-1988}. The twisted permeability tensor of the top AFMI slab is wrriten as $\mu^{'}=\mathcal{R}(\theta) \mu {\mathcal{R}}^{{T}}(\theta)$ with the rotation matrix $\mathcal{R}(\theta)$ along $z$ axis.
	
	Distinct from nonmagnetic crystals, the indefinite behavior in magnetic media can be controlled with a magnetic field, which affects the magnon-polaritons resonance frequency. If a magnetic field is applied as to generate spin canting, see Fig. \ref{fig1}(c)(left), the main effect is to shift the resonance $\omega_{\perp}$ to higher frequencies. On contrary, see Fig. \ref{fig1}(c)(right), irrespective of magnetic field, the resonance $\omega_{{\parallel}}$ basically unchanged. Therefore, the resonance and its features, such as the condition to obtain negative refraction where $\mu_{yy}\mu_{zz}<0$ and where $\mu_{yy}<0$ and $\mu_{zz}<0$, are also tuned to different frequencies\cite{Rair-2016,Rair-2019}. These hyperbolic behavior plays a significant role in NFRHT.
	
	During the numerical calculation, we adopt the parameters of manganese difluoride (MnF$_{2}$) which have been reported\cite{Remer-1986}, anisotropy force $B_A=0.787$ T, exchange force $B_E=55.0$ T, sublattice magnetization $M_s=6.0\times{10}^{5}$ A/m, permittivity $\varepsilon=5.5$ and gap distance $d=20$nm. At a temperature of 28 K, we have gyromagnetic ratio $\gamma/2\pi c =0.877$ cm${^{-1}}$/T and damping parameters $\Gamma/\omega_{0}=6.5 \times 10^{-4}$.
	
	\section{Results and discussion}
	{\it Symmetry analysis for near-field radiative heat transfer--} Before illustrating the main findings, we first give a basic symmetry analysis for MP-dominated radiative heat flux. As shown in Fig. \ref{fig1}(a), applied magnetic field can cause nonreciprocal surface electromagnetic mode on one AFMI slab due to time reversal symmetry breaking (\ding{172}$\rightarrow$\ding{175}). Regardless of the external magnetic field in \ding{173} and \ding{176}, the transmissions between two AFMI slab both have high degree of symmetry in $k_x-k_y$ space under inversion symmetry protected configuration due to the mismatch between different nonreciprocal MPs. When configurational inversion symmetry is breaking from \ding{176} to \ding{177} (twist angle $\theta=\pi$), we can find a strong nonreciprocal effect in the transmission plot due to the strong near-field coupling between nonreciprocal MPs at such configuration. Similar results are not found in the reciprocal-MPs dominated system under configurational symmetry breaking (\ding{174})\cite{Pajovic-2021}. After qualitative analysis of symmetry, Fig. \ref{fig1}(b) quantitatively shows that the NFRHT between the AFMI slabs can be severely controlled by the magnetic field and twist angle $\theta$. Primarily, the maximum value of HTC is about twice the minimum value, which respectively correspond to the light region and dark region in the phase image. Moreover, it is obvious that the presence of a magnetic field make the HTC change nonmonotonically and twist operation induces the thermal switch effect. We will demonstrate it in detail in the following
	discussions.
	
\begin{figure}
		\centering
		\includegraphics[width=\columnwidth]{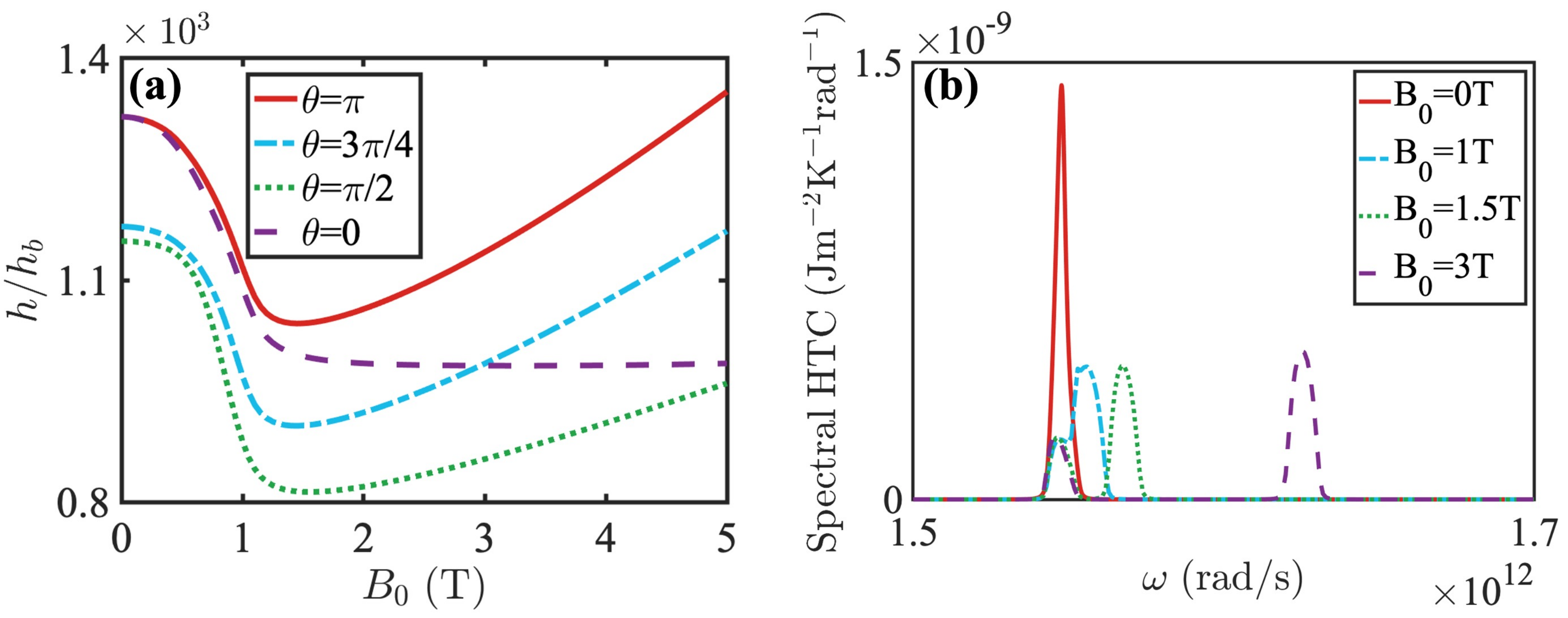} \\
		\caption{(a) The heat transfer coeﬀicient varying with magnetic field at different twist angle. (b) The heat transfer coeﬀicient spectral function with different magnetic field where the twist angle $\theta=\pi$. }
		\label{fig2}
	\end{figure}

	{\it Magnetic field dependence of NFRHT.--}
	In Fig. \ref{fig2}(a), we show the HTC as a function of the value of magnetic field for different twist angle. There are three salient features: (\romannumeral1) irrespective of the twist angle, the heat transfer decreases and then increases with the increase of magnetic field, (\romannumeral2) the maximum reduction of HTC up to $25\%$ can be found with magnetic field of about $1.5$ T, (\romannumeral3) in the interval where the magnetic field has positive feedback to the heat transfer, the enhancement effect of heat transfer decreases with the increase of twist angle, in particularly the HTC is nearly unchanged where the twist angle $\theta=0$. The strong modification of heat transfer due to the magnetic field is even more apparent in the spectral HTC. As one can see in Fig. \ref{fig2}(b), the magnetic field not only distorts and reduces the height of the peaks related to the surface waves, but it also generates two new peaks which eventually replace the initial peak. One of the peaks remained basically unchanged. On contrary, the other peak shifts to higher frequencies as the field increases. This additional peak appears at the resonance $\omega_{\perp}$ and its presence illustrates the high tunability that can be achieved.
	\begin{figure*}
		\centering
		\includegraphics[width=\textwidth]{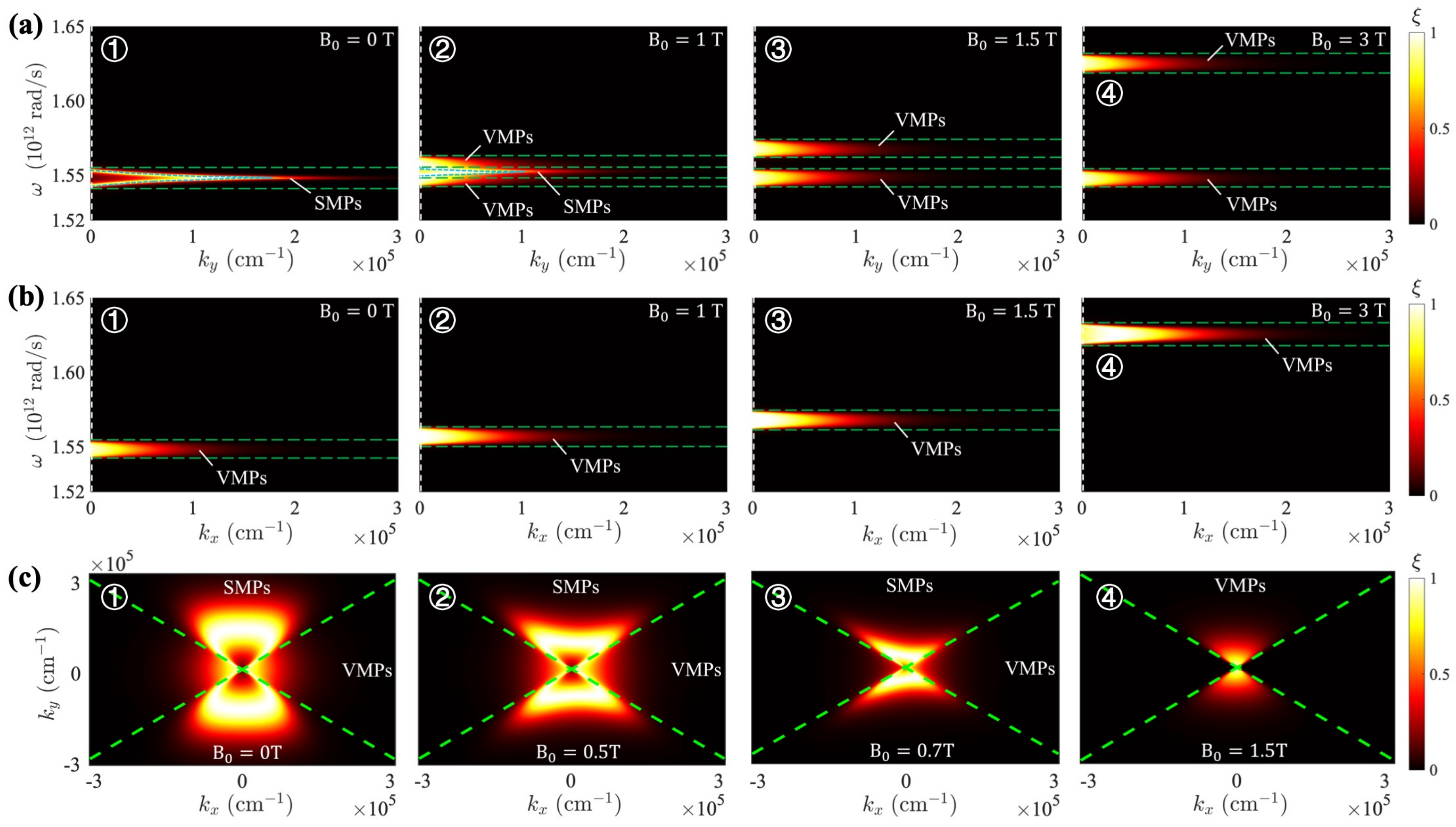} \\
		\caption{(a) Energy transmission coeﬀicients for wave vector along $y$-direction for various magnetic fields. (b) Energy transmission coeﬀicients for wave vector along $x$-direction for various magnetic fields. The horizontal dashed lines separate the regions where transmission is dominated by surface waves or propagating waves. The cyan dashed lines correspond to the dispersion relation of the SMPs and the white dashed lines represent light cone in the vacuum. (c) Energy transmission coefficient with the twist angle increasing at different magnetic fields. In (c), the frequency $\omega$ is fixed to $1.548\times10^{12}$rad/s and the green boundary lines for SMPs and VMPs are $k_{y}=\pm\sqrt{{-\mu_{xx}}/{\mu_{yy}}}k_x$. Note that in (a)-(c), twist angle $\theta=\pi$.}
		\label{fig3}
	\end{figure*}
	
	To shed more light on these results, it is convenient to examine the transmission coefficients in different directions of wave vector. We focus on exploring the $s$-polarized waves which can be shown to dominate the heat transfer. Fig. \ref{fig3}(a-b) presents corresponding energy transmission coefficients in $\omega-k$ space with different values of the magnetic field. On one hand, if the wave indicates along $y$-direction, the transmission maxima are located around a restricted area of $k$ and $\omega$ at low magnetic field, indicating that surface waves dominate the NFRHT. Fig. \ref{fig3}(a)\ding{172} also shows that the magnetic field can modify the dispersion relation of magnon polaritons. By increasing the magnitude of magnetic field,
	the channel of SMPs are progressively replaced by the channel of VMPs and the surface waves are restricted to the narrow reststrahlen band for higher magnetic field; see Fig. \ref{fig3}(a)\ding{173}. As shown in Fig. \ref{fig3}(a)\ding{174}-\ding{175}, at large magnetic field, the NFRHT is dominated by the channel of VMPs. It is obvious that the transmission corresponding to the low frequency band has little changes and the transmission corresponding to the high frequency band only has a frequency shift which is dependence of value of magnetic field. One the other hand, considering the wave indicate along the $x$-direction, Fig. \ref{fig3}(b)\ding{172}-\ding{175} shows that the VMPs throughout dominate the NFRHT and the frequency and cutoff wave vector $k$ corresponding to transmission maxima gradually increase with the increase of the magnetic field. To sum up, the modes which dominate the NFRHT change from SMPs into VMPs with the magnetic field increase.

	To explain the nature of above magnetic field induced modes, it is significant to notice that the off-diagonal elements of the permeability tensor (nonreciprocal term) plays little roles in radiative heat transfer ($\mu_{xz} \ll \mu_{xx,yy,zz} $). So we assume that the polarization conversion is irrelevant and the plates effectively behave as biaxial media where their permeability tensors are diagonal: $\hat{\mu}=$ diag[$\mu_{xx},\mu_{yy},\mu_{zz}$]. Within this approximation (biaxial approximation), it is easy to compute the dispersion relation of the SMPs: finding the singular value of denominator of the transmission coefficient, i.e.,
	\begin{align}\label{dispersion-SMPs}
		1-r_{ss}^{2}e^{-2|k_{z}|d}=0,
	\end{align}
	where reflection coefficient $r_{ss}=\frac{k_{z}{\mu}_{yy}-k_{z}^{'}}{k_{z}{\mu}_{yy}+k_{z}^{'}}$ and  $k_{z}^{'}$ is the $z$-component of wave vector in the AFMI. The cyan dashed lines in Fig.~\ref{fig3}(a-b) shows that the dispersion relation reproduces the maximum of the transmission in the frequency regions there surfaces waves are allowed ($\mu_{yy},\mu_{zz}<0$). The nature of VMPs can also be understood within the biaxial approximation. After matching the boundary conditions, the dispersion of the VMPs with different wave indicate directions can be written as below:
	\begin{align}
		\begin{split}
			&{\frac{k_{y}^{2}}{{\mu}_{zz}}+\frac{{k_{z}^{'}}^{2}}{{\mu}_{yy}}=\epsilon(\frac{\omega}{c})^2},\\
			&{{\frac{k_{x}^{2}}{{\mu}_{zz}}+\frac{{k_{z}^{'}}^{2}}{{\mu}_{xx}}=\epsilon(\frac{\omega}{c})^2}}.
		\end{split}
	\end{align}
	The dispersion of VMPs becomes hyperbolic when $\mu_{yy}\mu_{zz}<0$ or $\mu_{xx} \mu_{zz} <0$. As illustrated in Fig. \ref{fig3}(a)\ding{173}-\ding{175} and Fig. \ref{fig3}(b)\ding{172}-\ding{175}, the hyperbolic regions correspond exactly to the areas where the transmission reaches its maximum for a broad range of $k$ values. This fact shows that our AFMI plates effectively behave as hyperbolic materials.
	
	{\it Field-induced topological transition of magnon polaritons
		.--} More magnetic effects about SMPs and VMPs can be presented at $k_x-k_y$ space with fixed frequency.
	As shown in Fig.~\ref{fig3}(c), SMPs and VMPs in $k_x-k_y$ space are seperated by the green boundary lines, i.e., $k_{y}=\pm\sqrt{{-\mu_{xx}}/{\mu_{yy}}}k_x$ with the biaxial approximation. At $B_0=0$ (Fig. \ref{fig3}(c)\ding{172}), the whole systems is reciprocal and there is coexistence of $xy$-symmetrical ellipse SMPs and VMPs. With the increase of the magnetic field in $y$-direction, time-reversal symmetry is broken and two types of modes gradually show non-reciprocity at the $x$-direction. Next, we notice that the topological structure of SMPs in $k_x-k_y$ space can change from ellipse to hyperbolic with magnetic field increasing (Fig.~\ref{fig3}(c)\ding{173}-\ding{174}). And such field-induced topological transition can reduce the contributions of SMPs in NFRHT. Moreover, we also find that there is a transition from type I VMPs to type II VMPs with magnetic field increasing and there is only type II VMPs mode at large magnetic field (Fig.~\ref{fig3}(c)\ding{175}). Such field-induced transition effects can pave a more flexible way for thermal managements in nanoscale.
	
	\begin{figure}
		\centering
		\includegraphics[width=\columnwidth]{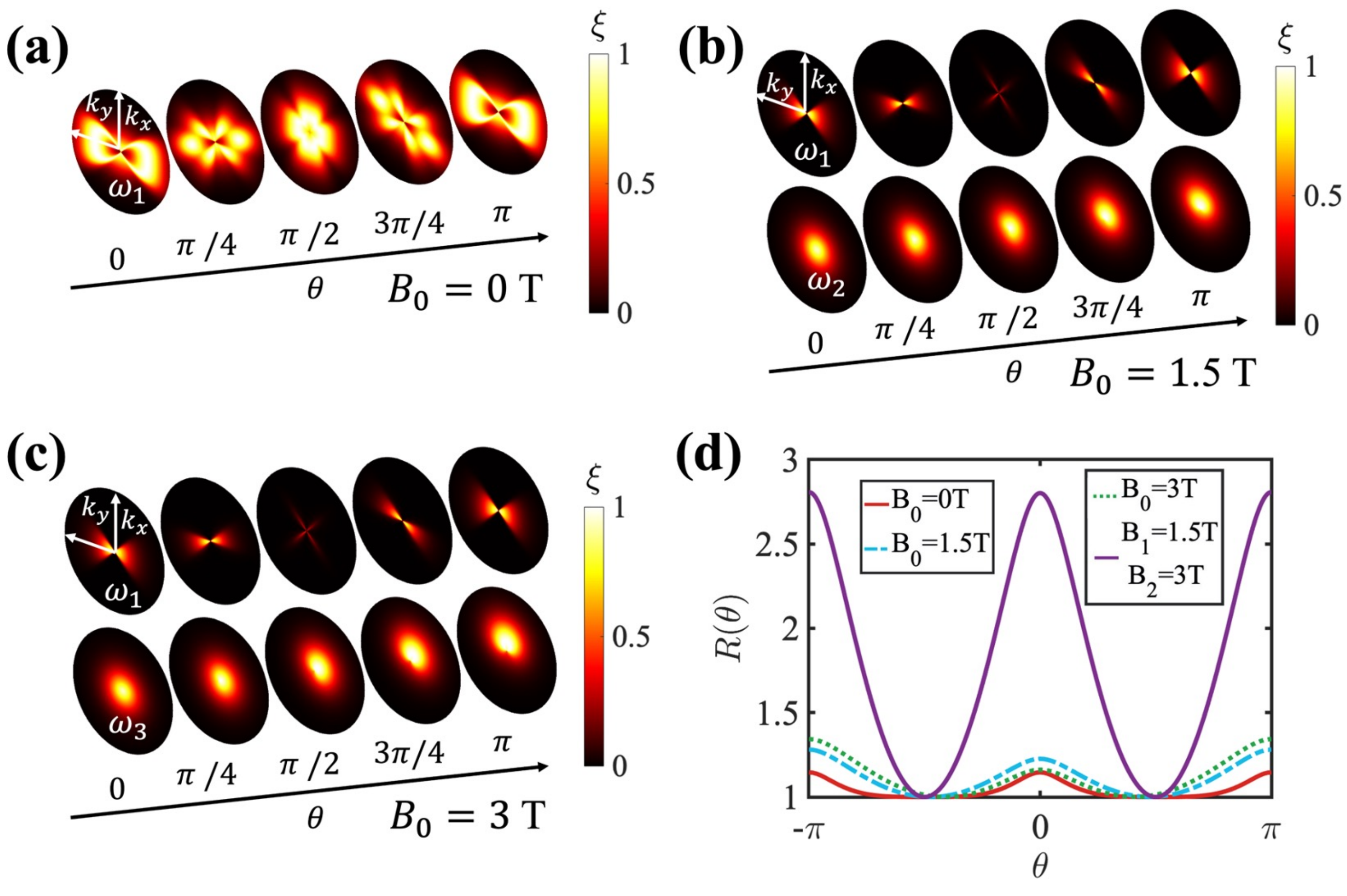} \\
		\caption{(a) Energy transmission coefficient with the twist angle increasing when $B_{0}=0$T. (b) Energy transmission coefficient with the twist angle increasing when $B_{0}=1.5$T. (c) Energy transmission coefficient with the twist angle increasing when $B_{0}=3$T. (d) The thermal switch ratio as a function of twist angle with different magnetic field. Particularly, the purple line correspnd to the case that applied magnetic fields on the two AFMI slabs are not equal. In (a)-(c), the frequency $\omega_{1}=1.548\times10^{12}$rad/s, $\omega_{2}=1.568\times10^{12}$rad/s and $\omega_{3}=1.625\times10^{12}$rad/s.}
		\label{fig4}
	\end{figure}
	
	{\it $\theta$-dependence of NFRHT and magnetic-field reinforced thermal switch.--} We analyse the influence of twist angle $\theta$ on heat transfer in Fig.~\ref{fig4}. At $B_0 = 0$ (Fig.~\ref{fig4}(a)), there is coexistence of SMPs and VMPs separated by the above metioned boundary line. With the increase of twist angle $\theta$, we find that there is hybridization between SMPs and VMPs and the boundary between two modes can become trivial at $\theta=\pi/2$. Such twist-induced hybridization provides a convenient way for thermal managements. Fig.~\ref{fig4}(b) and (c) shows the cooperative effects about magnetic field and twist angle $\theta$. We find that two field-induced VMPs have completely different twist effects: low-frequency VMPs can be almost suppressed at $\theta=\pi/2$; On the contrary, high-frequency VMPs only
	slightly change the shape in $k_x-k_y$ space. In addition, we notice that the high-frequency VMPs have a symmetrical structure at $\theta = 0$ and remain nearly unchanged as magnetic field increasing due to the whole system become symmetric again at $\theta=0$. Such effect corresponds to purple-dashed line in Fig.~\ref{fig2}(a). Above cooperative effects can construct a magnetic-field reinforced thermal switch. We define the thermal switch ratio $R(\theta)$ as below\cite{Jiebin-2021}:
	
	\begin{align}
		R(\theta)=h(\theta)/h_{min}.
	\end{align}
	Fig. \ref{fig4}(d) shows that the strength of magnetic field can only enhance thermal switch ratio in a relatively small range, i.e., from 1.1 at $B_1=B_2=0T$ to 1.2 at $B_1=B_2=1.5T$ or 1.25 at $B_1=B_2=3T$. But the thermal switch ratio can greatly boosted by the different configurations of applied magnetic fields: from 1.1 at $B_1=B_2=0T$ to 2.8 at $B_1=1.5 T,B_2=3 T$. Combining all panels of Fig. \ref{fig4}, we can see that such magnetic-field reinforced thermal switch effect comes from the mismatch of high-frequency VMPs at different applied magnetic fields. From the view of symmetry analysis, we can find that this is a cooperative effect between rotational symmetry breaking and external magnetic field (time reversal symmetry breaking). Based on that, we can theoretically control thermal switch ratio through external magnetic field in a wide range.
	
	\section{Conclusion}
	In summary, we show that the near-field radiative heat transfer between two hyperbolic antiferromagnets can be controlled by the external magnetic fields or twist angle $\theta$ between two plates. Under time reversal symmetry breaking, we find that the presence of a magnetic field make the radiative heat flux change nonmonotonically irrespective of twist angle, and the minimum value can be found with fields of about 1.5 T. From the spectral function of HTC, the heat flux contributed by SMPs is gradually replaced by VMPs with field increase. Then SMPs can take place striking topological transition with magnetic field increase and similar types of VMPs have also changed. Moreover, we find that twist induced thermal switch stronger with field increase due to nonzero off-diagonal elements dependence of nonreciprocity of VMPs. To harvest an ultrastrong twist induced thermal switch, we make an active choice for thermal transfer channel using different external magnetic fields, keeping the heat exchange channel sensitive to twist angle $\theta$ under configurational symmetry breaking. Finally, all the predictions of this work are amenable to measurements with the present experimental techniques, and we are convinced that the multiple open questions that this work raises will motivate many new theoretical and experimental studies of this subject.
	
	\begin{acknowledgments}
		The work is supported by the National Natural Science Foundation of China (No. 11935010) and the Opening Project of Shanghai Key Laboratory of Special Artificial Microstructure Materials and Technology.
	\end{acknowledgments}

	\bibliography{MP_ref}
\end{document}